# Matter-wave soliton control in optical lattices with topological dislocations

Yaroslav V. Kartashov and Lluis Torner

*ICFO-Institut de Ciencies Fotoniques, and Universitat Politecnica de Catalunya,
Mediterranean Technology Park, 08860 Castelldefels (Barcelona), Spain*

We address the concept of guiding and transporting matter-wave solitons along the channels of optical lattices made by interference patterns of beams containing topological wavefront dislocations. The local lattice distortions that occur around the dislocations cause solitons to move along reconfigurable paths, a phenomenon that may be used for controlled all-optical manipulation of Bose-Einstein condensates. Multiple dislocations form traps that can capture and hold moving solitons. We also show that suitable lattice topologies may be used to explore matter-wave soliton collisions and interactions in a confined reconfigurable environment.

Interactions between atoms forming Bose-Einstein condensates result in a rich variety of physical phenomena. Among them is the formation of different types of nonlinear self-sustained structures, or matter-wave solitons [1]. Various types of matter-wave solitons have been experimentally observed to date in geometries where the condensate confinement is achieved by external magnetic and optical traps. This includes dark [2], bright [3], and gap [4] solitons created in condensates with both repulsive and attractive inter-atomic interactions, as well as the recent landmark experimental observation of three-dimensional multiple soliton formation in condensates with inter-atomic attractive interactions hold in three-dimensional traps [5].

Bose-Einstein condensates can also be hold in optical lattices (for a recent overview see, e.g., Ref. [6]), a physical setting that affords a wealth of important new phenomena. In particular, lattices profoundly affect both the interactions between matter-wave solitons and their stability. A key advantage that makes optical lattices especially interesting is their intrinsic tunability. Namely, the lattice shape and its depth can be readily tuned by varying the propagation angles and the intensities of the lattice-creating light beams. This tunability, in turn, allows the manipulation of condensate clouds loaded into the lattice (see, e.g. Refs. [7-15]). The concept of reconfigurable optical lattices finds also important applications in nonlinear optics, where lattices can be induced optically in suitable, highly anisotropic nonlinear materials (see, e.g., Refs. [16-18]). The simplest periodic optical lattice can be constructed by several interfering plane waves. Such periodic lattices support various types of soliton states, whose properties are dictated by the specific topology of the lattice creating waves. In particular, formation of soliton trains consisting of condensate droplets localized in the vicinity of the lattice maxima is possible [19-21]. In addition to sets of plane waves, other types of nondiffracting light beams, such as Bessel beams, can be used for lattice induction [22]. The symmetry of such beams result in new opportunities for soliton manipulation, such as rotary soliton motion (experimentally demonstrated in [23]) and rotary soliton interactions, as well as reconfigurable soliton wires, arrays, and matrices for soliton routing and transport [24]. The accessible possibilities are restricted to the symmetry of the lattice.

A qualitatively different type of optical lattices may be constructed with optical landscapes made by the interference pattern of regular light waves and singular beams that contain topological wavefront dislocations. On physical grounds, one expects soliton behavior in the vicinity of dislocations to change drastically thus resulting in new possibilities for soliton control. For example, we predicted recently that in lattices made by the interference of tilted plane waves and beams carrying screw phase dislocations, or optical vortices, solitons experience attractive and repulsive forces by the dislocations, which may result in transverse displacements of solitons that were initially *at rest* [25]. Such phenomena suggest specific potential applications to *dynamical matter-wave soliton manipulation*, which we address in details in this paper. More specifically, here we consider, for the first time to our knowledge, the collisions of *moving* solitons with dislocations as well as complex phase-dependent interactions of *several moving* solitons in the vicinity of dislocations, which could find applications for *controllable condensate transport* with topologically complex reconfigurable lattices.

We study in detail how the local distortion of the guiding lattice channels around the dislocations drags solitons along the corresponding paths, thus affording a new way to transport and to sort simultaneously matter-wave solitons that can be brought together to collide in the vicinity of the dislocations. We also find that suitable arrangements of multiple dislocations can form different soliton traps which are able to capture and hold solitons moving along the lattice channels. We show that the concept may be used to explore specific types of soliton interactions and interferences, because those are strongly affected by the geometry of the optical lattice.

The phenomena we study here arise naturally in pancake-shaped Bose-Einstein condensates hold in suitable lattices made with spatial light modulators, and they might hold too in nonlinear optics provided that the corresponding lattices can be made stable and robust. Therefore, for the sake of generality, we perform our study in the framework of a generic model of matter-wave and light evolution in a cubic nonlinear medium. We thus address the evolution of a Bose-Einstein condensate hold in an optical lattice, described by the nonlinear Schrödinger equation for the dimensionless mean-field wave function $q$:

$$i\frac{\partial q}{\partial \xi} = -\frac{1}{2}\left(\frac{\partial^2 q}{\partial \eta^2} + \frac{\partial^2 q}{\partial \zeta^2}\right) - q|q|^2 - pR(\eta,\zeta)q. \qquad (1)$$

Here the transverse coordinates $\eta, \zeta$ are normalized to the characteristic transverse scale of the condensate $a_0$; variable $\xi$ stands for time in units of $\tau = ma_0^2/\hbar$, with $m$ being the mass of the atoms and $\hbar$ the Planck's constant; parameter $p$ is proportional to the lattice depth $E_0$ expressed in units of recoil energy $E_{\text{rec}} = \hbar^2/ma_0^2$. Lattice depths of the order $E_0 \leq 22 E_{\text{rec}}$ ($p \leq 22$) have already been achieved [7,8]. The function $R(\eta,\zeta)$ describes the transverse lattice profile. The sign and the strength of mean-field cubic nonlinearity is determined by the s-wave scattering length $a_s$ that can be changed by the external magnetic field via Feshbach resonance. Further we consider attractive interactions with $a_s < 0$. In lithium condensates with characteristic transverse scale $a_0 = 5\,\mu\text{m}$ the dimensionless time $\xi = 1$ corresponds to 2.5 ms of actual evolution. Equation (1) admits several conserved quantities, including the soliton norm $U$ and the Hamiltonian $H$:

$$U = \iint_{-\infty}^{\infty} |q|^2 \, d\eta d\zeta,$$
$$H = \frac{1}{2} \iint_{-\infty}^{\infty} (|\partial q / \partial \eta|^2 + |\partial q / \partial \zeta|^2 - 2pR|q|^2 - |q|^4) d\eta d\zeta. \tag{2}$$

Here we assume that the optical lattice features the intensity of the interference pattern of a tilted plane wave and a beam carrying one or several screw wavefront dislocations, or optical vortices. In the case of a wave that carries a single screw phase dislocation, one has $R(\eta,\zeta) = |\exp(i\alpha\eta) + \exp(im\phi + i\phi_0)|^2$, where $\alpha$ is the propagation angle of the plane wave with respect to the $\eta$ axis, $\phi$ is the azimuthal angle, $m$ is the winding number or topological charge of the vortex, and $\phi_0$ stands for the orientation of the vortex origin in the transverse plane. In the case of multiple vortices the last term in the above expression for $R(\eta,\zeta)$ is to be replaced by the product of several functions describing vortices with charges $m_k$, positions $\eta_k, \zeta_k$, and orientations $\phi_k$. Optical lattices of this type can be generated, e.g., by using spatial light modulators [26,27], similar to those that have been successfully used for optical confinement of Bose-Einstein condensates [28].

Lattices produced by the interference of plane waves and waves with several nested vortices are not stationary, hence they slowly distort upon propagation. Nevertheless, they can be implemented in pancake-shaped, or disk-shaped, Bose-Einstein condensates, which are elongated in the transverse $(\eta,\zeta)$ plane but strongly confined in the direction of light propagation ($\rho$ axis), so that lattice-creating waves remain effectively constant inside the whole condensate. Such a strong confinement of the condensate can be achieved, e.g., in asymmetric magnetic harmonic traps with the ratio of oscillation frequencies $\nu_\rho/\nu_{\eta,\zeta} \sim 100$ (where $\nu_\eta = \nu_\zeta$) or in tight optical traps (see Refs. [29] for details of the experimental realization of lower-dimensional condensates). Thus, the thickness of a lithium condensate with a ground-state scattering length $a_s \approx -1.45$ nm [30] trapped in a potential with $\nu_\rho = 150$ Hz, can be $\sim 3\,\mu$m, that is much smaller than the characteristic length of the lattice distortion. We always set values of parameters in such way that the magnitudes of the linear and nonlinear terms in Eq. (1) are comparable. Thus, we let $\alpha = 4$ (which sets characteristic transverse scale of lattice in $\eta$ direction to $2\pi a_0/\alpha \approx 7.8\,\mu$m) and let $p = 1$. Finally, we can neglect the impact of additional trapping potentials in Eq. (1) because of the difference in the transverse scales of the parabolic potential and the optical lattice $R(\eta,\zeta)$.

Some representative examples of lattices produced by interference of a tilted plane wave and waves with nested vortices with different topological charges are depicted in Fig. 1. The presence of clearly pronounced lattice channels where matter-wave solitons can travel is apparent. Far from the dislocation, the lattice profile is similar to that of a quasi-one-dimensional lattice produced by the interference of the two plane waves $\exp(\pm i\alpha\eta)$. In contrast, because of the screw phase dislocation carried by one of the lattice-creating waves, several lattice channels fuse or vanish in the vicinity of the fork located at the position of the topological dislocation. This is accompanied by a local distortion of the neighboring lattice channels. The higher the winding number $m$ of the screw dislocation, the stronger the lattice distortion.

Far from the dislocation fork, the lattice channels can support stable solitons, whose maxima coincide with local maxima of the function $R(\eta,\zeta)$. We found families of such stationary solutions numerically in the form $q(\eta,\zeta,\xi) = w(\eta,\zeta)\exp(ib\xi)$, with $w(\eta,\zeta)$ being a real function, and $-b$ being the chemical potential. When launched towards the fork with large enough initial velocity $\alpha_\zeta$ (this can be achieved with standard phase-imprinting techniques [1]), solitons move along the channel where they are initially loaded. As a result, the position of the soliton center shifts by $m\pi/\alpha$ along the $\eta$ axis when solitons pass around a single dislocation fork (Fig. 1). Notice that when a soliton is launched into one of the channels that merge together in the fork, it gets either destroyed or it bounces back when it hits the dislocation fork. Such phenomenon makes possible the parallel transport of several solitons simultaneously launched into a lattice with several dislocations featuring multiple open and closed paths. This is illustrated in the bottom right plot of Fig. 1. The important idea behind this concept is the possibility to add or to remove vortices in the lattice-creating wave, hence creating new soliton paths along which several solitons can be brought together to collide in the vicinity of dislocation.

Suitable combinations of dislocations can form various lattice traps. The simplest trap is formed when two vortices nested in the lattice-creating wave feature opposite topological charges $m = \pm 1$ and are separated by the distance $\delta\zeta$ along the vertical direction. Depending on the mutual orientation of the vortices, the two channels of the resulting lattice can either fuse first and split in a second dislocation (we term this geometry a positive trap) or part of the lattice channel located between the dislocations can disappear (we term this geometry a negative trap). In what follows, we consider positive traps as those illustrated in Fig. 2(a). Notice that the lattice channels close to the one forming the trap are almost undistorted and therefore solitons launched into these channels at $\zeta < 0$ are readily transported to the region $\zeta > 0$. In contrast, solitons launched into one of the channels forming the trap can be either bounced back or captured by the trap. In the latter case, captured solitons oscillate inside the trap until a new, stationary *trap mode* is reached. We have found that soliton capturing by the lattice trap is only possible when the input soliton velocity $\alpha_\zeta$ exceeds a critical value, $\alpha_{\mathrm{cr}}$. Our calculations revealed that the critical velocity growths monotonically with the lattice depth and with the initial soliton norm (this is shown in Figs. 2(b) and 2(c)). The possibility of soliton capturing by lattice traps enriches the set of operations accessible with distorted lattices. For example, lattice traps can also be produced by sets of higher-order, oppositely charged vortices. In this case, at fixed $U$ and $p$ the critical velocity is found to rapidly decrease with increase of vortex charges $m$. Thus, at $U = 4.5$ and $p = 1$ one has $\alpha_{\mathrm{cr}}^{m=1} \approx 0.47$, while $\alpha_{\mathrm{cr}}^{m=3} \approx 0.25$.

The optical lattice landscapes that we address in here offer a unique setting to generate new types of matter-wave soliton interactions in a confined environment. We explored briefly the interactions between solitons guided by the lattice channels, and the outcome of the interaction can be strongly altered by the presence of the lattice dislocation. For illustrative purposes, here we concentrate in the case of two identical solitons with a phase difference $\delta\phi$ located in neighboring lattice channels that are launched simultaneously toward the trap with a given velocity, $\alpha_\zeta$. Numerical simulations show that the output strongly depends on the phase difference between solitons in new and non-trivial ways. In particular, attraction between in-phase solitons can cause penetration of both solitons inside the trap despite the fact that the initial velocity is well below the critical value (in this case $\alpha_{\mathrm{cr}} \approx 0.47$). This phenomenon is shown in Fig. 3(a). In contrast, accurate simulations reveal that two out-of-phase solitons cannot penetrate inside the trap and hence simultaneously bounce back even for velocities greatly exceeding the critical one. This phenomenon is depicted in Fig. 3(b). In the case of arbitrary phase difference between the input solitons, a oscillating state is formed with $U$ oscillating periodically between the involved lattice channels. Depending on the exact value of the initial phase difference, a single output soliton can be eventually formed and bounced back into one of the guiding channels after collision with the lattice trap. This phenomenon is illustrated in Figs. 3(c) and 3(d).

A variety of additional interactions are accessible with the topologically distorted lattices, by varying the parameters of the lattice-creating beams (e.g., the number, loci, and winding number of nested topological screw dislocations) and of the input soliton states (e.g., soliton norms and velocities). A comprehensive investigation of such interactions falls beyond the scope of this paper, but a few illustrative examples follow. For example, increasing the winding number of the topological screw dislocation nested in the lattice-creating waves results in a significant enhancement of the corresponding trap width. Thus, we found that out-of-phase solitons that collide in the vicinity of such wide channels may penetrate into the traps, despite of the repulsive forces acting between them; solitons might even escape from the trap provided that the input soliton velocities are high enough. This phenomenon is illustrated in Fig. 4(a).

Simulations reveal that the outcome of the collision of several solitons in the vicinity of the lattice dislocation depends strongly on the number of colliding solitons and on their relative phase differences. For example, launching additional solitons toward the dislocation from the region $\zeta > 0$ completely changes the outcome of the collision depicted in Fig. 4(a). When one additional soliton is launched at $\zeta > 0$ (it is in-phase with soliton launched into left-hand-side channel and out-of-phase with soliton launched into right-hand-side channel at $\zeta < 0$) three solitons emerge upon the collision process, with one of them being captured by the trap, another one being bounced back into one of the channels emerging from the trap, and only one soliton penetrates into the region $\zeta > 0$. An illustrative example is shown in Fig. 4(b).

Interestingly, collisions of several solitons arriving to the trap from both the upper (located at $\zeta > 0$) and the lower (located at $\zeta < 0$) channels may result in the formation of higher-order nonlinear trap modes, which might be termed *multipole-mode trap solitons*. Dipole trap solitons might be formed when two out-of-phase input solitons with initial velocities substantially exceeding a critical value collide inside the trap, as depicted in Fig. 4(c) and 4(d). Our calculations show that dipole-mode soliton form in this case for different choices of channels where the input solitons are launched.

Similarly, we verified that triple-mode trap solitons can be formed upon collision of three single solitons with appropriately chosen velocities and phases.

We thus conclude stressing that we have addressed in detail the concept of matter-wave soliton control in optical lattices containing topological fork dislocations. Such lattices can be created in pancake-shape Bose-Einstein condensates with spatial light modulators; they feature complex guiding channels where solitons can be routed and let interact. The concept is intended to hold for different optical landscapes produced by the interference of light waves carrying screw phase dislocations, including Laguerre-Gaussian modes. In this paper we pose the idea for matter-wave soliton states, but the concepts might hold as well for optical solitons, provided that stable suitable lattices can be inducted or technologically fabricated in nonlinear crystals in a robust way.

This work has been partially supported by the Government of Spain through grant TEC2005-07815/MIC and the Ramon-y-Cajal program.

# Figure captions

Figure 1. Snapshot images showing controllable shift of soliton that moves in the lattice channel in the vicinity of the dislocation created by vortices with charges 1, 3 and 5. Parallel soliton transport in the complex lattice with several dislocations created by unit-charge vortices. The initial velocity of soliton is $\alpha_\eta = 0$, $\alpha_\zeta = 2$, while its initial norm $U = 4.5$. Snapshot images are taken with step $\delta\xi = 2$.

Figure 2. (a) Dynamics of soliton capture by the positive lattice trap. Initial and final distributions of modulus of wave-function are superimposed. Initial soliton velocity $\alpha_\eta = 0$, $\alpha_\zeta = 0.8$, its norm $U = 4.5$, and trap length $\delta\zeta = \pi$. Critical angle versus lattice depth at $U = 4.5$ (c) and versus soliton norm at $p = 1$.

Figure 3. Collisions of solitons launched into neighboring lattice channels in the vicinity of positive trap. (a) $\alpha_\zeta = 0.3$, $\delta\phi = 0$. (b) $\alpha_\zeta = 0.7$, $\delta\phi = \pi$. (c) $\alpha_\zeta = 0.42$, $\delta\phi = \pi/2$. (d) $\alpha_\zeta = 0.42$, $\delta\phi = -\pi/2$. In all cases initial norm of each soliton $U = 4.5$ and velocity $\alpha_\eta = 0$. Arrows show direction of motion for input and output solitons.

Figure 4. Plots (a) and (b) show collision of solitons in the vicinity of positive trap produced by vortices with topological charges $m = \pm 2$. Plot (a) shows passage of out-of-phase solitons trough the trap for high initial velocities $\alpha_\zeta = 1.5$. Plot (b) shows complex three-soliton collision for $|\alpha_\zeta| = 1.5$, when soliton launched from the region $\zeta > 0$ is in-phase with one of two solitons launched from the region $\zeta < 0$. Plots (c) and (d) show excitation of second trap mode by two out-of-phase solitons with $|\alpha_\zeta| = 0.8$ launched into different channels of distorted lattice. In all cases initial norm of each soliton $U = 4.5$ and velocity $\alpha_\eta = 0$. Trap length $\delta\zeta = \pi$. Arrows show direction of motion for input and output solitons.

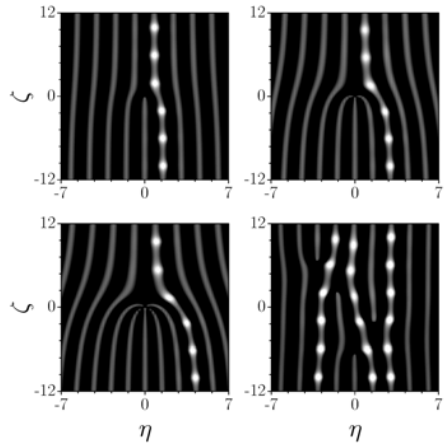

Figure 1.  Snapshot images showing controllable shift of soliton that moves in the lattice channel in the vicinity of the dislocation created by vortices with charges 1, 3 and 5. Parallel soliton transport in the complex lattice with several dislocations created by unit-charge vortices. The initial velocity of soliton is $\alpha_\eta = 0$, $\alpha_\zeta = 2$, while its initial norm $U = 4.5$. Snapshot images are taken with step $\delta\xi = 2$.

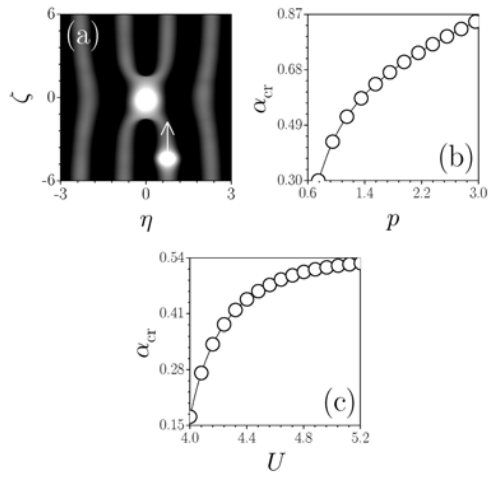

Figure 2. (a) Dynamics of soliton capture by the positive lattice trap. Initial and final distributions of modulus of wave-function are superimposed. Initial soliton velocity $\alpha_\eta = 0$, $\alpha_\zeta = 0.8$, its norm $U = 4.5$, and trap length $\delta\zeta = \pi$. Critical angle versus lattice depth at $U = 4.5$ (c) and versus soliton norm at $p = 1$.

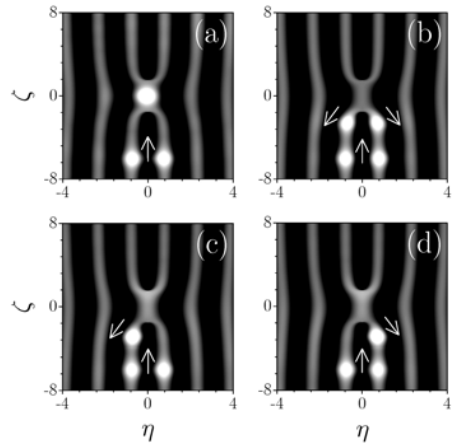

Figure 3.  Collisions of solitons launched into neighboring lattice channels in the vicinity of positive trap. (a) $\alpha_\zeta = 0.3$, $\delta\phi = 0$. (b) $\alpha_\zeta = 0.7$, $\delta\phi = \pi$. (c) $\alpha_\zeta = 0.42$, $\delta\phi = \pi/2$. (d) $\alpha_\zeta = 0.42$, $\delta\phi = -\pi/2$. In all cases initial norm of each soliton $U = 4.5$ and velocity $\alpha_\eta = 0$. Arrows show direction of motion for input and output solitons.

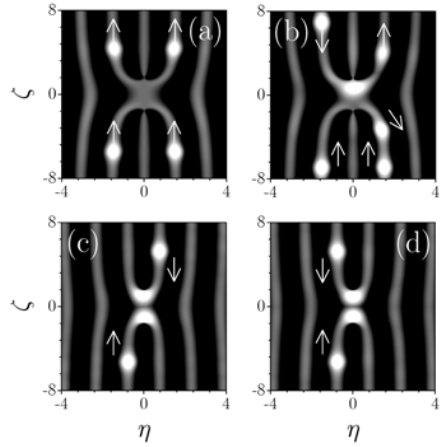

Figure 4. Plots (a) and (b) show collision of solitons in the vicinity of positive trap produced by vortices with topological charges $m = \pm 2$. Plot (a) shows passage of out-of-phase solitons trough the trap for high initial velocities $\alpha_\zeta = 1.5$. Plot (b) shows complex three-soliton collision for $|\alpha_\zeta| = 1.5$, when soliton launched from the region $\zeta > 0$ is in-phase with one of two solitons launched from the region $\zeta < 0$. Plots (c) and (d) show excitation of second trap mode by two out-of-phase solitons with $|\alpha_\zeta| = 0.8$ launched into different channels of distorted lattice. In all cases initial norm of each soliton $U = 4.5$ and velocity $\alpha_\eta = 0$. Trap length $\delta\zeta = \pi$. Arrows show direction of motion for input and output solitons.